# Si nano-VINe: Electrical Percolation in Quantum-Confined nc-Si:SiO$_2$ Systems

Serim Ilday[1,2,†], Hande Ustunel[3], F. Ömer Ilday[4], Daniele Toffoli[5], Gizem Nogay[2,3], David Friedrich[6], René Hübner[6], Bernd Schmidt[6], Karl-Heinz Heinig[6], Rasit Turan[2,3]

[1]Department of Micro and Nanotechnology, Middle East Technical University, 06800, Ankara, Turkey

[2]Center for Solar Energy Research and Applications (GÜNAM), Middle East Technical University, 06800, Ankara, Turkey

[3]Department of Physics, Middle East Technical University, 06800, Ankara, Turkey

[4]Department of Physics, Bilkent University, 06800, Ankara, Turkey

[5]Department of Chemistry, Middle East Technical University, 06800, Ankara, Turkey

[6]Institute of Ion Beam Physics & Materials Research, Helmholtz-Zentrum Dresden-Rossendorf, D-01328 Dresden, Germany

† To whom correspondence should be addressed: serim.ilday@metu.edu.tr

**Abstract**

Nanostructures, especially of silicon, are of paramount importance for new generation solar cell applications. The key material requirements for solar cells are good electrical conductivity, confinement of excitons, and a tunable band-gap that allows for tandem arrangements to absorb a maximally large portion of the solar spectrum. However, to date, existing Si-based nanostructures have addressed these requirements individually. The difficulty is that these requirements are entangled in complex ways, so an effort to improve one can lead to a major trade-offs in the others. In this study, we report on a novel silicon nanostructure, where the key advance is that we have managed to satisfy all the abovementioned requirements simultaneously: We demonstrate that the excitons are confined, the bandgap is tunable, and the electrons can freely travel along the nc-Si network.

**Introduction**

The search for an optimal material for use in photovoltaic devices has been ongoing for more than three decades[1-7]. A consensus has been reached on the essential features of the optimal material: it should be of low cost, have good electrical conductivity and a tunable band-gap, which allows for tandem arrangements to absorb a maximally large portion of the solar spectrum[8-12]. Silicon nanostructures such as quantum-dots[6], -wells[10], -wires[7,13], and porous silicon[1,14] constitute promising candidates, but they achieve these intricately coupled properties individually and an unguided optimization of one can have detrimental effects on the other. A small number of device demonstrations have been realized, albeit with inadequate efficiencies[15-17]. Here, we introduce Si nano-VINe (**V**ertically **I**nterconnected **Ne**twork of silicon) as a vertically oriented, electrically percolated and quantum-confined silicon nanocrystal (nc-Si) network with a tunable bandgap, embedded inside a $SiO_x$ matrix. Good electrical conduction is achieved by promoting directional growth for the nc-Si network using ballistic deposition[18-21], which lowers the critical density of Si required to achieve a highly connected network[22,23]. Lowering of the Si concentration assists in preserving quantum-confinement, which is simultaneously achieved through fine-tuning of chemical bonding dynamics of the material system.

Percolation is the spontaneous emergence of macroscopic connectedness in a network of randomly distributed connections, when a critical threshold in the density of the connections is exceeded. Percolation is a ubiquitous phenomenon, observed in countless systems, such as neural networks[24], biological systems[25], the world-wide-web[26], and social networks[27]. Similarly, a highly connected electrical network can emerge among randomly distributed nc-Si in a dielectric host matrix and this percolated network can efficiently transport the photo-excited carriers, fulfilling the requirement of good electrical conductivity. However, nominally, the high concentrations of Si required to achieve percolation cause effective diameters of the nanocrystals to swell beyond the critical exciton Bohr radius (~5 nm for Si), which ruins quantum confinement (QC). Attempts to control percolation of nc-Si:SiO$_x$ systems by tuning stoichiometry are well known[28-35], but it was not clear *a priori* whether QC could be preserved, as required for a tunable bandgap, while simultaneously forming a percolated network that can efficiently transport the photoelectrons. In other words, the challenge has been to engineer the material system such that there is a range of Si concentrations, for which the percolation threshold is exceeded, while the exciton Bohr radius remains below the critical level. Guided by theory, we have developed an optimized magnetron-sputtered thin-film deposition and post-annealing process that meets these requirements simultaneously.

**Experimental**

*Thin Film Deposition and Characterization*

nc-Si:SiO$_x$ thin films were deposited onto p-type, 1-5 Ω·cm, (100)-oriented Si wafers and quartz substrates via magnetron-sputtering technique using Vaksis, NanoD-100 PVD system. 3"-diameter 99.995%-pure SiO$_2$ and Si targets were co-sputtered with applied RF ($P_{SiO2}$ = 180 W) and DC powers (10<$P_{Si}$<190 W), respectively. Depositions were performed at room temperature and 99.99% pure Ar was used as the process gas, where base pressure was ~2x10$^{-6}$ Torr. Post-annealing was performed in a high-temperature furnace for 1h under constant flow of N$_2$ gas at 1100°C. In order to reveal the vine-like morphology in the magnetron-sputtered and post-annealed nc-Si:SiO$_x$ thin films, energy-filtered transmission electron microscopy (EFTEM) imaging was performed using an image-corrected FEI Titan 80-300 microscope operating at an accelerating voltage of 300 kV and equipped with a Gatan Imaging Filter 863. Classical cross-sectional TEM specimens were prepared by sawing, grinding, dimpling, and final Ar$^+$ ion-milling. The same analysis results were obtained for TEM lamellae prepared by in situ lift-out using an NVision 40 CrossBeam device (Zeiss). Prior to each TEM analysis, the specimen mounted in a double-tilt analytical holder was placed for 45 s into a Model 1020 Plasma Cleaner (Fischione) to remove organic contamination. Stoichiometry, *x* in SiO$_x$, were derived through Rutherford backscattering spectrometry (RBS) elemental analyses and confirmed via X-ray photoelectron spectroscopy (XPS) elemental analyses. RBS analyses were performed using 1.7 MeV He$^+$ ions accelerated by a Van-de-Graaff generator. After normal incidence backscattered ions were detected at 20° exit angle, where their charge was chosen to be 10 µC for good S/N ratio. SimNMR software (version 6.03) was used to fit and analyze the elemental composition. Atomic fractions and areal densities obtained from RBS and XPS measurements were used both in the numerical model and in the molecular dynamics simulations. Conductivity analyses were carried through both vertical and lateral IV measurements. p-type, 0.01-0.05 Ω·cm, (100) oriented Si wafers were used for vertical IV measurements whereas quartz substrates were used for lateral IV measurements. For vertical IV analyses the current flow was measured between thermally evaporated aluminum point contacts on film surface and Al contact evaporated onto the back surface of the Si wafer. For lateral IV analyses ~1 µm thick coplanar Al contacts were evaporated with 1.5 cm spacing onto film surface on a quartz substrate. Current flow between the thin-film and Si substrate for vertical IV analyses and current flow between coplanar electrodes for lateral IV analyzes was measured via Keitley 2440 ampermeter and Hewlett – Packard 4140B pikoampermeters, respectively. Newport Oriel Sol3A Class AAA solar simulator at 0.9 A.M. was used for the light-illuminated I-V measurements. XPS analyses were performed using PHI-5000 VersaProbe equipment with monochromatic Al $K_\alpha$ excitation as the X-ray source. XPS depth profile analyses were conducted using 99.99% pure Ar ion bombardments, where the energy of incoming ions was kept at 2 keV. Fourier-transform infrared spectroscopy (FTIR) analyses were performed using Brucker Equinox 55 IR Spectrometer in absorbance mode. The vibrational spectra of the thin films were recorded between 300 and 4000 $cm^{-1}$ at a resolution of 4 $cm^{-1}$ at a 40° angle of incidence. Photoluminescence (PL) measurements were taken under Nd:YAG laser excitation (532 nm line) at room temperature. PL setup used to record the PL excitation spectra comprises an Nd:YAG laser, Hamamatsu C7041 detector head, Oriel instruments MS257 monochromator and a series of lenses and mirrors. The spectra are recorded in the range of 550-1100 nm. Broad PL signal between ~650-1100 nm was observed for $0.6<x<1.5$ each centered around ~860 nm where the signal intensity differs with $x$ value. Optical bandgap values were extracted from Tauc plots calculated through transmittance, reflectance, and absorbance spectra taken in the 180-1200 nm wavelength range by using a double beam scanning Cary 100 UV-Vis spectrophotometer equipped with Czerny Turner monochromator.

*Numerical Model of Ballistic Deposition*

Simulations of the ballistic growth are based on layer-by-layer deposition of Si and $SiO_2$ atoms to sites in a 3D grid. The probability of any site being occupied by a Si atom, $p_0$, depends on the overall deposition rate of Si. Given the cold substrate, the mobility of a Si atom is heavily reduced. Thus, when considering the affinity of Si atoms to seek out and stick to each other, we model only the nearest-neighbors, which are described by two probabilities, $p_1$ (for the site directly below) and $p_2$ (for the immediate neighboring sites), respectively. The relative strength of $p_1$ over $p_2$ determines the tendency towards vertical growth (anisotropic growth) over branching of laterally (isotropic growth). The value of $p_0$ is directly fixed by experiment as it determines the Si concentration value of $x$. The values of $p_1 = 10$ and $p_2 = 100$ are chosen to reproduce the 3D morphology of the structure observed from the TEM images as closely as possible. Although the values of $p_1$ and $p_2$ are not precisely fixed by experiment, we found out that the critical $x$ value to achieve percolation in the vertical or horizontal direction does not change appreciably by the specific values of $p_1$ and $p_2$, as long as they are chosen to resemble the experimentally observed structure. When the simulation code is run, each site is calculated to be either a Si site, flagged with a value of 1 or a non-Si site, flagged with a value of 0, layer by layer until the full thin film is built up. After the simulation of the deposition process is finished, we calculate which Si sites are connected through each other to conducting layers at the top and bottom of the grid, representing electrodes sandwiching the thin film. We use a recursive algorithm to determine the connections, starting from each site in the top and the bottom layers. For each site, the code takes steps in all six directions, unless the stepped into site is outside of the boundaries of the grid. If a stepped-into site is a Si site, it is flagged with a value of 2 and further steps are taken in all 6 directions starting from that site. If the site is a non-Si site, then its flag is left unchanged and no other steps are taken from that site. When the simulation code exhausts sites from which to take further steps, the final state is a grid with each site labeled as 0, 1 or 2, representing a non-Si site, a disconnected Si site and an electrically connected Si site. The ratio of number of total flag 1 and flag 2 sites to flag 0 sites corresponds to $x/(1+x)$, where $x$. Next, we calculate the ratio of number of flag 2 sites to flag 1 sites, corresponding to the ratio of connected Si sites to disconnected Si sites. Since the photoelectrons generated at disconnected sites are thermalized and do not contribute to current formation, high-efficiency device operation necessitates the connected sites to dominate over disconnected sites by a ratio of ~5 or more. The simulation is very strongly recursive, and the number of sites to visit scales with order of $6^n$, where $n$ is the total number of sites in the grid. The simulations are repeated for several times for each value of $x$ to get an average value. Fortunately, the connectedness ratio is found to vary very little, independent of the grid size (beyond a minimum size of, e.g., 50x50x50) and from run to run. In other words, the results are robust. Despite the simplicity of our model, its agreement is good enough to serve as guide to the experiments, which is not surprising given the universality of such growth models, rendering the conclusions to be invariant against small variations in the details.

*Molecular Dynamics Simulations*

Molecular dynamics (MD) simulations were conducted on homogeneous Si:SiO$_x$ systems for 0.5<$x$<1.5 at the experimental annealing temperature. The atomic scale of this approach allows for a local map of Si-O bonding and determination of individual Si partial charges for various $x$ values and all histograms are drawn as a percentage of the total number of Si atoms in the system. Si-Si, O-O, and Si-O atomic interactions are modeled using the charge-optimized many-body (COMB) potential[38,39] and the open-source molecular program suite LAMMPS is used for the calculations[40]. The MD simulations runs were conducted in a 2.6 x 2.6 x 2.6 nm$^3$ cubic simulation cell with periodic boundary conditions. A total of 1000 Si and O atoms were initially regularly positioned on a cubic grid, with a predetermined O:Si ratio of $x$ ($x$ = 0.5, 0.7, 1.0, 1.2 and 1.5). The distribution of Si and O atoms on the grid were chosen at random from a uniform probability distribution. A time-step of 0.5 fs was chosen for the integration of equations of motion. An initial 20000-step long equilibration stage conducted in the NVE ensemble was followed by a 200000-step long MD simulation using the Langevin thermostat at a nominal temperature of 1100 °C. Charges of each Si and O atoms are treated as a dynamical variable in the COMB potential are updated at every step of the MD simulation. The Si-Si, O-O and Si-O interaction thus include both covalent effects and charge transfer. The Coulombic charge assigned to each of the atoms in the simulation is recalculated at every integration step.

**Results and Discussion**

Our approach is based on optimizing "global" structure for electrical percolation and "local" structure for QC preservation separately, while tying them up through the chemical bonding dynamics. The desired global network properties are achieved by promoting preferentially vertical growth of the nc-Si network. The preferentially vertical growth allows us to reach percolation, hence good electrical conductivity first along the vertical direction[22,23], which matters most for a solar cell. To this end, we use a cold substrate kept at room temperature, with which energetic particles continuously collide during deposition. Since the substrate is much colder than the incoming particles, the deposited particles lose their kinetic energies rapidly upon hitting the surface, which forces them to settle at a minimum energy position within a few nanometers of the collision point, instead of seeking out the global minimum over large portions of the surface. This stochastic deposition results in the formation of a randomly connected nc-Si network. The restricted mobility due to the cold surface increases the probability of Si particles being stacked up in the vertical direction, relative to horizontal. This small difference in probabilities drives the emergence of a slightly anisotropic, preferentially vertically growing nc-Si network, i.e., the Si nano-VINe, akin to vine climbing a wall. We developed a simple numerical model of the ballistic growth mechanism to guide our thinking as we experimentally optimized the ballistic deposition growth parameters.

Preferentially vertical growth of the nc-Si network is confirmed in a ~330 nm-thick film through energy-filtered transmission electron microscopy (EFTEM) analysis, in particular valence-band plasmon energy-loss imaging at $E_{loss}$ = 17 eV (Fig.1a). The pseudo-colored superposition of the Si (green) and SiO$_x$ (red) plasmon EFTEM image of the same field of view is shown in Fig.1b. Our model reproduces the experimentally observed transition from a disconnected archipelago of QC nc-Si (pre-nano-VINe region for $x$>1.3, where $x$ is the O:Si ratio of the system) (Fig.1c), to a percolated network (nano-VINe region for 1.3>$x$>0.9) (Fig.1d), which is yet sparse enough to preserve QC, finally to a virtually completely connected but overcrowded network (post-nano-VINe region for $x$<0.9) (Fig.1e), for which the QC is irrevocably lost. These results are generally consistent with ballistic deposition theory, which itself falls under the Kardar-Parisi-Zhang (KPZ) universality class[18-21]. Likewise, the lateral connectedness of the Si nano-VINe structure can be observed in the zero-loss-filtered high-resolution TEM (HRTEM) image in Fig.1f, which only shows Bragg-oriented nc-Si, as well as in the pseudo-colored superposition of the Si (green) and SiO$_x$ (red) plasmon EFTEM image in Fig.1g. Both images are from exactly the same field of view with a size of approximately 25 nm x 25 nm and are obtained for the same sample as shown in Fig.1a and b. As can be seen from the figures that the structure is connected both in lateral and in vertical directions while preserving the average nc-Si sizes at ~2 nm, which is below the critical Bohr radius for Si.

We have performed current (I) - voltage (V) measurements between thin-film and Si substrate (Fig.2a) and determined the conductivity as a function of $x$ for vertical contacts (Fig.2b). Similarly, we have conducted I-V measurements between two coplanar aluminum electrodes evaporated onto film surface on a quartz substrate (Fig.2c) and determined the conductivity for horizontal contacts (Fig.2d). It is seen that the electrical conduction in vertical direction increases abruptly at $x = 1.3$, which corresponds to a Si concentration well below the nominal percolation threshold ($x \sim 1$) of Si in $SiO_x$ matrix. As to be expected, conductivity between horizontal contacts shoots up at a higher Si concentration, at $x \sim 1.1$, which serves as an indirect, though strong confirmation of the preferentially vertical connectedness of the Si nano-VINe structure. The magnitude of the lateral conductivity is much lower than that of the vertical conductivity since the distance between the two Al electrodes was 1.5 cm, which is an excessively large distance for charge carriers with short lifetimes. Importantly, it is demonstrated that the structure is conducting both vertically and laterally with relatively good conductivity, indicating percolated current flow without relying on tunneling current. We also show that there is photoelectric effect in nano-VINe and post-nano-VINe regions (insets of Fig.2b and 2d) as seen from the increased conductivity when measurements are taken under light-illuminated conditions.

The second leg of our approach addresses the local structure with the aim of QC preservation without disrupting the conductivity-oriented global properties as discussed above. Locally, our deposition process creates a medium, composed of nominally unstable Si suboxides ($Si^{1+}$, $Si^{2+}$, and $Si^{3+}$), in addition to stable Si ($Si^{0+}$) and $SiO_2$ ($Si^{4+}$) species. These suboxides are heterophase structures, which restrict the mobility and spatial distribution of the Si and O atoms during ballistic deposition, thus affording us a level of control over Si agglomeration. Normally, when sufficient energy is applied to the system (generally by post-annealing at high-temperatures), Si suboxides break down and turn into stable forms of either Si or $SiO_2$ through a disproportionation reaction[36]. If too many suboxides turn into Si, they inflate the effective nanocrystal size, which eventually destroys the QC effect. If too many suboxides turn into $SiO_2$, they surround and isolate the Si quantum dots (QD), which have to rely on tunneling currents for conduction[11,23,33]. The key issue becomes, then, to obtain nominally unstable Si suboxides and to prevent them from stabilizing completely upon addition of external energy. The ballistic deposition used to promote vertical growth also serves as a tool for controlling the local chemistry: During thin film growth, new atoms/clusters continue to collide with the cold surface where they braid a random network into which the Si and O atoms are stuck. Even when a carefully calibrated amount of external energy is introduced into the system via post-annealing, the atoms are not able to move freely to achieve complete phase separation, but they are energized enough to rearrange locally. This limitation helps preserve the suboxides and prohibits further expansion of the effective nc-Si diameter, while stabilizing the material. Consequently, a large number of small nanocrystals are obtained, which preserves QC at relatively high concentration levels.

We performed detailed X-ray photoelectron spectroscopy (XPS) and Fourier-transform infrared spectroscopy (FTIR) analyses on post-annealed thin films to demonstrate that a nominally unstable suboxides are stabilized and the structure does not reach full phase separation after addition of external energy (Fig.3a-c). XPS analyses show that both post-nano-VINe (Si-rich) and pre-nano-VINe ($SiO_2$-rich) regions are almost completely phase separated, the former consisting mostly of $Si^{0+}$, whereas the latter consists mostly of $Si^{4+}$ (Fig.3a and c). However, in the nano-VINe region, all suboxides are present as well as $Si^{0+}$ and $Si^{4+}$ (Fig.3a), which clearly demonstrates that complete phase separation is prevented and the stabilized suboxides force the structure to arrange itself in a way that underlies the Si nano-VINe structure. In Fig.3b, we plot the integrated peak areas of stoichiometric anti-symmetric Si-O-Si stretching vibrations signal and the signal coming from $SiO_x$ using FTIR analysis. These results show that the signal due to $SiO_2$ decreases in the nano-VINe region, while the signal due to the nominally unstable $SiO_x$ increases strongly. We have observed that the key features and findings of XPS and FTIR chemical bonding dynamics analyses are supported by the molecular dynamics simulations (Fig.3d and e). Fig.3d shows histogram of Si coordination number defined as the number of O atoms within a certain radius (results are presented for a value of 2 Å) surrounding each Si atom and therefore an indicator of the Si oxidation state, which shifts from low to high values as $x$ increases. This is in agreement with the observed trend of increasing Si partial charge reported in the histograms in Fig.3e, where only the partial charges of the Si atoms are considered. In other words, a value of 10 on the vertical axis and a corresponding bin of 1.25-1.50 on the horizontal axis indicate that 10% of the Si atoms have a partial charge of between +1.25e and +1.50e.

In the Si nano-VINe structure, unlike QDs, excitons are not confined in all three dimensions (Fig.1g), resulting in partial loss of confinement for each nanocrystal along the spatial direction it is connected to another nanocrystal. This is consistent with the photoluminescence (PL) analyses. Fig.4a shows that PL signal intensity of isolated Si

QDs (pre-nano-VINe region with nc-Si diameter of ~2 nm) is pretty strong, whereas it abruptly decreases in the Si nano-VINe region due to the connecting nanocrystals. However, thanks to the nano-VINe structure being randomly connected, QC is still preserved on average, as evidenced from the less intense but still present PL emission (Fig.4a) and the spectral blueshift with respect to varying nanocrystal size (inset of Fig.4a). The preservation of QC is further confirmed from the size-dependent optical bandgap of Si nano-VINe structure (Fig.4b), where we have used Tauc plots to extract the bandgap values. We show that the bandgap decreases from ~4 eV for the pre-nano-VINe region to ~1.65 eV for the post-nano-VINe region. For the Si nano-VINe region, the bandgap ranges between ~$1.75 < E_g < ~3.1$ eV, which covers the visible spectrum, where half of the solar power lies. These results are also consistent with valence-band plasmon energy-loss TEM images that show increasing of Si nanocrystal sizes with increasing Si concentration for the nano-VINe region (Fig.4c-f). Thus, we see that the Si nano-VINe material system sacrifices QC at a level that it still works well for PV applications, while achieving good electrical conduction.

**Conclusion**

To conclude, we propose a novel nanoengineered structure, Si nano-VINe, as a powerful candidate for use in PV devices, in which the excitons are confined, bandgap is tunable, and the electrons can freely travel along the nc-Si network. The Si nano-VINe structure is expected to serve well towards band alignment for a final all-Si nano-VINe device framework based on a tandem solar cell structure with its configurable bandgaps. We report on an interdisciplinary and theory-guided methodology that untangles the mutual dependencies of electrical conductivity, exciton confinement, and tunable band-gap requirements. Our approach is based on separate optimization of the local and global properties, where the partial decoupling is rendered possible by the universality of the global network properties. This approach can potentially be used to nanoengineer other material systems. While the immediate application is to PV devices, the Si nano-VINe structure could serve as an anode material for rechargeable lithium batteries[37], as well as numerous optoelectronic devices that require electrically conducting, yet quantum-confined materials[11,12].

**Acknowledgements**

This work was supported by the Scientific and Technological Research Council of Turkey (TÜBİTAK) and German Federal Ministry of Education and Research (BMBF) with Grant No. 109R037. The authors thank to İlker Yıldız, and Seçkin Öztürk for their contributions to XPS and early HRTEM analyses, respectively.

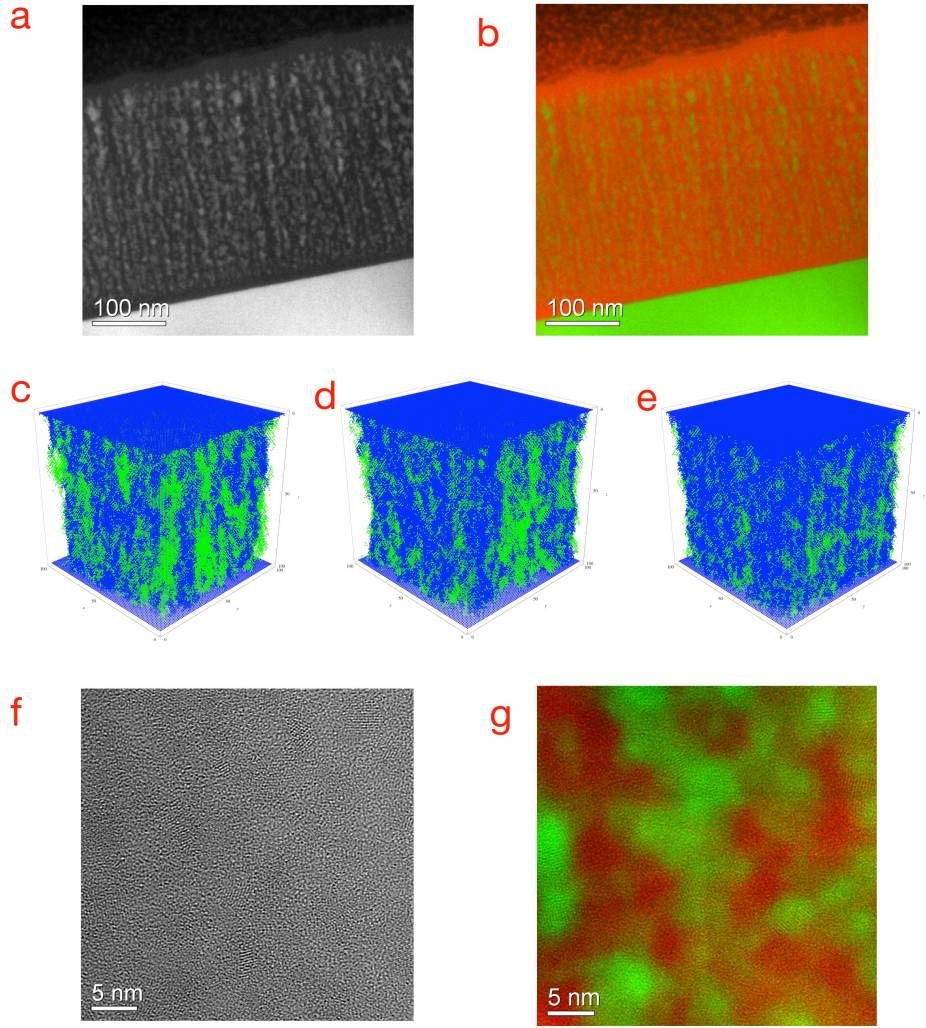

**Figure 1** Cross-sectional TEM images showing preferential growth of Si nano-VINe structure in vertical direction **(a)** Si (bright) plasmon EFTEM image. **(b)** Superposition of the Si (green) and the $SiO_x$ (red) plasmon EFTEM images. Numerical modeling showing universal features of the growth dynamics (blue areas are isolated Si clusters, red areas are connected Si clusters) of **(c)** pre-nano-VINe region for $x = 1.46$, **(d)** nano-VINe region for $x = 1.07$, **(e)** post-nano-VINe region for $x = 0.51$. Cross-sectional TEM images of Si nano-VINe structure with an average nc-Si size of ~2 nm **(f)** Zero-loss-filtered HRTEM image (dark areas are Si, bright areas are $SiO_x$). **(g)** Superposed Si (green) and $SiO_x$ (red) plasmon EFTEM image of the same field of view as in (f).

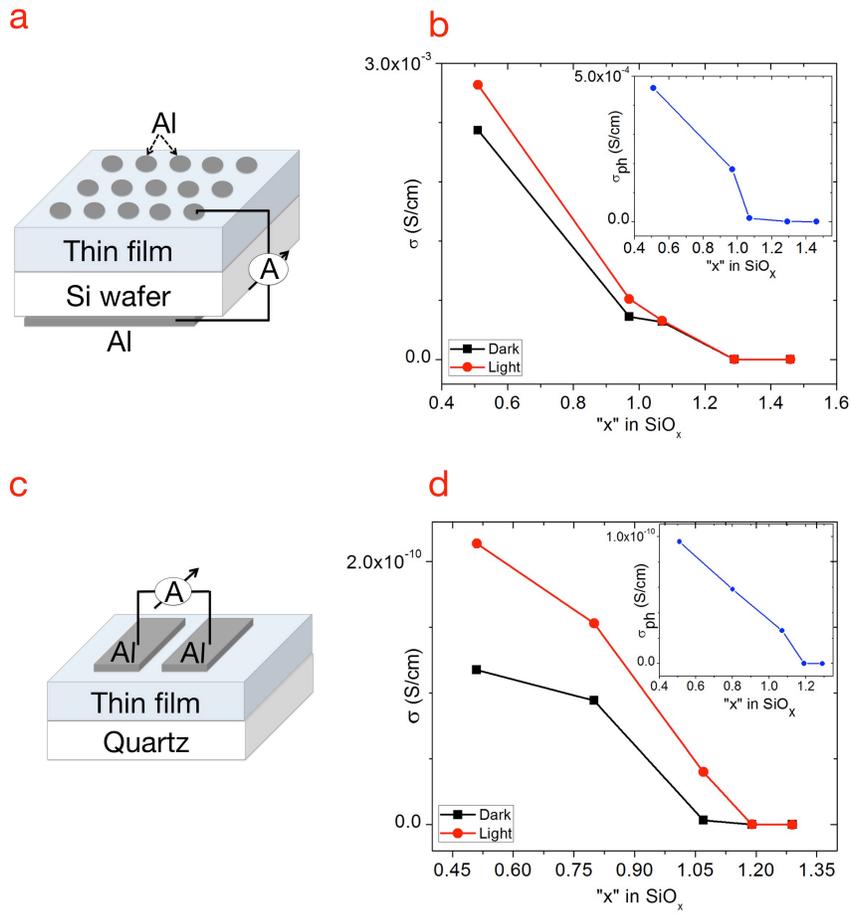

**Figure 2** Electrical conductivity in vertical direction **(a)** Cartoon of the device used to conduct I-V measurements. **(b)** Graph showing an abrupt increase in the vertical electrical conductivity at $x = 1.3$ (black line represents dark conductivity, red line represents conductivity under light-illuminated conditions), (inset) is the graph showing the photoelectric effect in the vertical direction. Electrical conductivity in lateral direction **(c)** Cartoon of the device used to conduct I-V measurements. **(d)** Graph showing an abrupt lateral electrical conductivity increase when $x\sim1.1$ (black line represents dark conductivity, red line represents conductivity under light-illuminated conditions), (inset) is the graph showing photoelectric effect in lateral direction.

Figure 3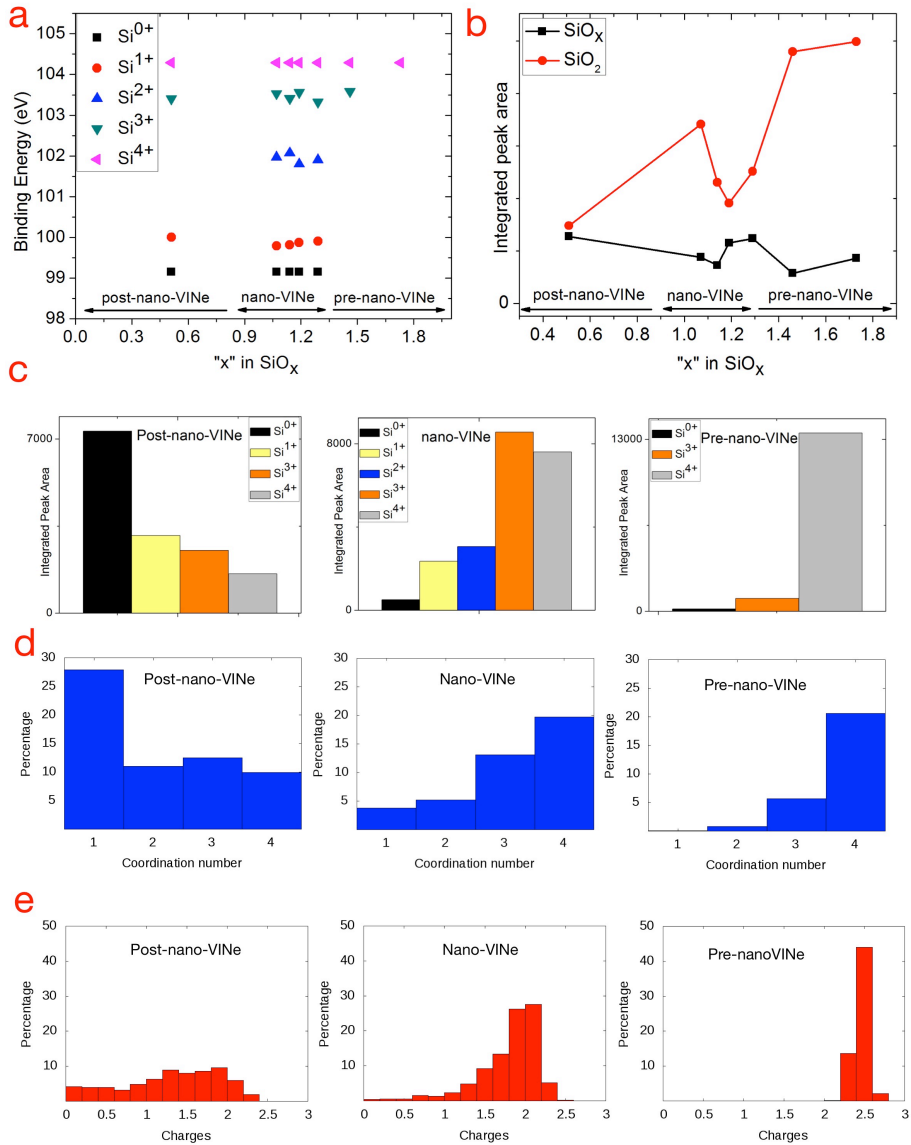

**Figure 3** XPS and FTIR analyses show that nominally unstable medium is achieved and suboxides are stabilized **(a)** Graph showing XPS signal positions of $Si^{n+}$ ($n$ = 0, 1, 2, 3, 4) oxidation states and which $Si^{n+}$ species are exist in pre-nano-VINe, nano-VINe, and post-nano-VINe regions. **(b)** Graph showing the relative contributions of stoichiometric, stable, $SiO_2$ signal to the nominally unstable, $SiO_x$ signal to FTIR spectra for pre-nano-VINe, nano-VINe, and post-nano-VINe regions. **(c)** Graph showing relative contributions

of $Si^{n+}$ signals to XPS spectra for post-nano-VINe (for $x$ = 0.51), nano-VINe (for $x$ = 1.19), and pre-nano-VINe (for $x$ = 1.46) regions. Molecular dynamics simulations show **(d)** Si coordination number histograms for post-nano-VINe ($x$ = 0.5), nano-VINe ($x$ = 1), and pre-nano-VINe ($x$ = 1.5) regions. **(e)** Si charge histograms for post-nano-VINe, nano-VINe, and pre-nano-VINe regions.

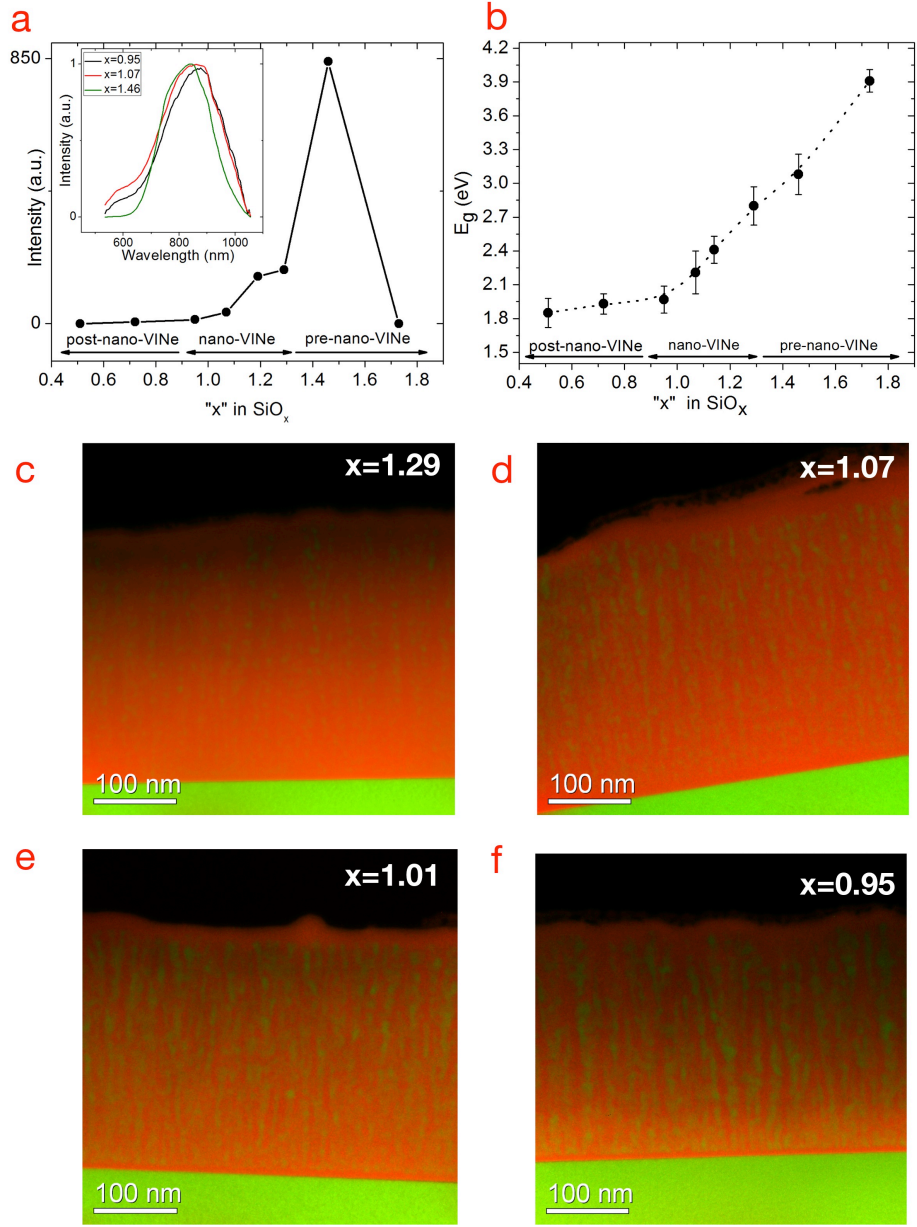

**Figure 4** Optical analyses of Si nano-VINe structure **(a)** Graph showing a highly intense PL signal for pre-nano-VINe region, where there are QDs, and an abrupt PL signal intensity decrease for nano-VINe region, and almost no PL signal for post-nano-VINe region, (inset) Graph showing size-dependent spectral blueshift of PL emission. **(b)** Graph showing size-dependent change in optical bandgap from pre-nano-VINe to nano-VINe and post-nano-VINe regions. Superposed cross-sectional Si (green) and $SiO_x$ (red) plasmon EFTEM images of Si nano-VINe structure for various $x$ values **(c)** $x$ = 1.29, **(d)** $x$ = 1.07, **(e)** $x$ = 1.01, and **(f)** $x$ = 0.95.